\documentclass[journal]{IEEEtran}

\usepackage[english]{babel}

\usepackage[letterpaper,top=2cm,bottom=2cm,left=2cm,right=2cm,marginparwidth=1.75cm]{geometry}

\usepackage{subcaption}
\usepackage{amsmath}
\usepackage{graphicx}
\usepackage[colorlinks=true, allcolors=blue]{hyperref}

\usepackage{epstopdf}

\usepackage{varioref}
\usepackage{float}
\usepackage{comment}

\usepackage{xcolor}
\usepackage[linesnumbered,ruled,vlined]{algorithm2e}
\usepackage{amsmath}
\usepackage{caption}
\usepackage{graphicx}
\usepackage{hyperref}

\SetCommentSty{mycommfont}

\SetKwInput{KwInput}{Input}                
\SetKwInput{KwOutput}{Output}              

\usepackage{flushend}
\usepackage{tabularx,booktabs}
\newcolumntype{C}{>{\centering\arraybackslash}X} 
\newcolumntype{L}{>{\raggedright\arraybackslash}X} 
\usepackage[noadjust]{cite}

\usepackage{wasysym}
\usepackage{lipsum} 
\usepackage{setspace}
\usepackage{newtxtext,newtxmath}
\begin{document}

\title{Efficient Fog Node Placement using Nature-Inspired Metaheuristic for IoT Applications}

\author{Abdenacer Naouri, Nabil Abdelkader Nouri, Sahraoui Dhelim, Amar Khelloufi, Abdelkarim Ben Sada and Huansheng Ning
	\thanks{Abdenacer Naouri, Huansheng Ning, Amar Khelloufi, Abdelkarim Ben Sada are with the University of Science and Technology Beijing, Beijing 100083, China}
	\thanks{Sahraoui Dhelim is with the School of Computer Science, University College Dublin, Ireland.} 
    \thanks{Nabil Abdelkader Nouri is with the Department of Mathematics and Computer Science, University of Djelfa, Djelfa, Algeria} 
    
 \thanks{Corresponding author: Sahraoui Dhelim (sahraoui.dhelim@ucd.ie).}
}

\maketitle

\begin{abstract}
Managing the explosion of data from the edge to the cloud requires intelligent supervision such as fog node deployments, which is an essential task to assess network operability. To ensure network operability, the deployment process must be carried out effectively in terms of two main factors: connectivity and coverage. The network connectivity is based on fog node deployment which determines the physical topology of the network while the coverage determines the network accessibility. Both have a significant impact on network performance and guarantee the network QoS. Determining an optimum fog node deployment method that minimizes cost, reduces computation and communication overhead, and provides a high degree of network connection coverage is extremely hard. Therefore, maximizing coverage as well as preserving network connectivity is a non-trivial problem. In this paper, we proposed a fog deployment algorithm that can effectively connect the fog nodes and cover all edge devices. Firstly, we formulate fog deployment as an instance of multi-objective optimization problems with a large search space. Then, we leverage Marine Predator Algorithm (MPA) to tackle the deployment problem and prove that MPA is well-suited for fog node deployment due to its rapid convergence and low computational complexity compared to other population-based algorithms. Finally, we evaluate the proposed algorithm on a different benchmark of generated instances with various fog scenario configurations. The experimental results demonstrate that our proposed algorithm is capable of providing very promising results when compared to state-of-the-art methods for determining an optimal deployment of fog nodes. 
\end{abstract}

\section{Introduction}
With the recent advancement of the Internet of Things (IoT), we have witnessed the emergence of many new applications in different domains such as health care, surveillance, and smart cities. IoT systems can be realized using different computing and communication architectures, where each architecture can provide solutions for data processing, storing, and analyzing according to the available resources \cite{cui2021survey}. Fog computing and edge computing are considered integral parts of the IoT systems and are proposed to cover the cloud limitation issues arising from the enormous IoT sensor offloaded requests, such as latency response time, which may take a long time to answer users' requests \cite{asi}. Both fog computing and edge computing are regarded as an extension of cloud computing that can provide distributed computing, storage, and networking functions close to the IoT sensors at the edge network to serve the IoT applications needs which support low latency, real-time response, location awareness, devices mobility, scalability, heterogeneity, transient storage, high-speed data dissemination, decentralized computation, and security ~\cite{naouri2021novel}. The enormous resulting requests from IoT sensors and edge devices could lead to a data explosion through the network, thereby an efficient network deployment is required.  In most Ad-hoc networks, the deployment of computing nodes is accomplished either in a pre-planned or in an ad-hoc method. The pre-planned deployment method is used when the deployment field access is limited and the deployment cost is not expensive. While the ad-hoc manner is used when the deployment field is large, costly, and consists of many computing nodes~\cite{trust2vec}. 

The deployment strategy is constructed on the basis of several considerations such as the IoT sensor and edge device location, system functionality, and the environments. Also, it may differ from one area to another according to its nature, whereas in an open area, the deployment tries to cover a wider area than the closed area such as inside a building. There are plenty of deployment possibilities for user coverage and connectivity needs, for example in critical situations such as natural disasters, oil rigs, mines, battlefield surveillance, high-speed mobile video gaming, or in public transport \cite{khelloufi2020social}. Finding an optimal deployment for a wide area is a challenge that could suffer from user density and the unregulated deployment surface compared to small areas. Because the performance of the entire system may be impacted by the deployment of computing nodes which is considered a crucial concern \cite{wang2020fog}. Therefore, in this work, we concentrate on providing an optimal fog node placement within the edge network to serve the boundary edge devices and IoT sensors in a specific area. 

In order to deploy the fog computing nodes effectively, we need a full understanding of the relation between the network topology and node density, subjected to different factors such as location and density \cite{maiti2019effective}. Hence, some issues should be addressed to enhance the network performance including connectivity, coverage, reliability, accessibility, etc. If the fog nodes are installed without taking into account restricting factors of the real region of interest and the underlying topology, it can cause low network coverage and connectivity. The node deployment problem has been investigated with different networks and in different environments. It remains a very challenging problem that has been proven to be NP-Hard \cite{kumar2022survey}. In order to address these issues, metaheuristic techniques have been developed, although they typically only provide optimal local solutions \cite{natesha2022meta}. 

In comparison to other approaches, metaheuristic methods have experienced the greatest success and development in addressing a variety of real-world optimization issues. The computing scenario considered for this work consists of IoT sensors and edge devices located at the edge of the network and fog nodes located within the fog layer. As the problem showed its NP-Hard, we propose a MPO population-based metaheuristic algorithm to solve the fog node deployments within the edge network devices in an efficient way and evaluate the network coverage and connectivity effect on the performance of the network. Our contributions in this study are summarized  as follows:

\begin{itemize}

\item To effectively connect the fog nodes and cover all edge nodes, we propose a multi-objective aggregate objective function aimed at maximizing edge device coverage and fog node connectivity at the same time, ensuring that all devices and fog nodes are connected and covered. This will effectively enhance the network performance and guarantee the network QoS.
	
\item We provide an MPA metaheuristic nature-inspired optimization algorithm, where the objective function is set to achieve maximum coverage and connectivity for the network. We analyze how various parameters affect the network's performance and determine the most effective ways to optimize.
	
\item For seamless computation, the proposed algorithm is implemented and tested in various network settings. The results show a significant improvement in coverage and connectivity compared to other methods.

\end{itemize}

The rest of the  paper is organized as follows: Section \ref{sec.2}  reviews existing research of fog node deployment. Section \ref{sec.3}  describes the system modelling and problem formulation.  Section \ref{sec.4} details the proposed metaherictic algorithm for fog node deployment. Section \ref{sec.5} presents the  evaluation details and  experiment results. We conclude  the paper and  outline future research directions in Section \ref{sec.6}.

\section{Related work}
\label{sec.2}
Today networks require precise computing node placement within network architecture, for reliable measurement and efficient data transmission. For this purpose,  node deployment procedures are frequently mentioned as one of the most crucial methods; Although several researchers have been conducted to address this issue, the existing solutions must be improved upon or replaced with new ones on a regular basis due to their limitations. To achieve this, it is possible to employ optimization algorithms to find the optimum node locations that meet the specified criteria. 

Heuristics and metaheuristic methods have been used to solve the Fog Node deployment Problem (FND) with different objectives. Authors in~\cite{Yoon2013AnEG} initially addressed the coverage problem in WSN and demonstrated that it is NP-hard. They concentrated on the issue of improving the computing node's coverage for a given number of sensors. In~\cite{Yu2022} authors suggested an improved Artificial Bee Colony method to optimize the lifetime of a two-tiered wireless sensor network through optimum relay node deployment. The dimension of the problem is first integrated into the candidate discovery formula, and the local search is changed based on the fit of the issue and the number of iterations which assists in balancing the algorithm's exploration and exploitation capacity. In~\cite{Cong2015}, the authors proposed a particle swarm optimization (PSO) algorithm to optimize the node network deployment and improve the adaptive ability of the network, however, the algorithm has the disadvantage of falling into the local optimum. In \cite{Deng2019}, the authors presumed that the network's topology is in hexagonal form and suggested an improved virtual spring force algorithm (VSFA) to deploy the network nodes within. This essentially decreases the area of vulnerability, but the proposed strategy has not been evaluated in a complex setting It has just been tested under perfect circumstances. As in~\cite{Deng2019}, authors in~\cite{Qin2018} performed their evaluation with an ideal deployment environment presuming that nodes are homogeneous, where an enhanced fish swarm algorithm (AIFS) was adopted to optimize the node's deployment by targeting the coverage rate, which greatly increases the network coverage area and reduces the energy usage,  Authors in ~\cite{Wang2018WirelessSN}~\cite{XU2018268} concentrated on improving energy utilization, energy balance amount, and network coverage, whereas in~\cite{Wang2018WirelessSN} they combined three optimization objectives into a single objective in a linear weighted fashion, then used the whale group algorithm (WGA) to optimize the single objective function. Although the approach is simple, it consumes too long time for computation. In~\cite{XU2018268} authors only consider the deployment of homogeneous nodes and the presence of obstacles and they use two versions of a multi-objective evolutionary algorithm one based on decomposition and the other to jointly optimize the model.  

Authors on~\cite{Akusta2022} address fault tolerance and connectivity issues in WSNs. They concentrated their efforts on connection restoration and introduced a simple and fast algorithm for motion k-connectivity reconstruction, which separates nodes into critical and non-critical clusters based on their failure to lower k. When a crucial node fails, the algorithm grabs and transfers the non-critical nodes. The authors of~\cite{Khalilpour2021} presented methods for estimating connectivity in IoT-enabled wireless sensor networks. They introduce a simple and efficient method (PINC) for movement-based k-connectivity restoration that splits nodes into critical and non-critical groups. When a crucial node fails, the PINC algorithm takes over and transfers the non-critical nodes. This approach transfers a non-critical node with the fewest of movement cost to the position of the failed mote.


Unlike previous works which focused on placing nodes in a discrete grid area, which limits the positions of nodes. Our approach gives us high flexibility to place fog nodes within a continuous region of interest. In addition, previous approaches optimized network connectivity and user coverage in two hierarchical ways, while in our work, we considered a multi-objective aggregate function to optimize both of them at the same time, because user coverage may also be crucial in some practical services. Some experimental results will be provided to show the advantage of simultaneous optimization. Our approach gives us high flexibility to place fog nodes within a continuous region of interest. In addition, previous approaches such as in ~\cite{s19010032,Zeng2018CostEffectiveES} optimized network connectivity and user coverage in hierarchical ways, while in our work, we considered a multi-objective aggregate function to optimize both of them at the same time, because user coverage may also be crucial in some practical services.

\section{System assumptions and problem model formulation}
\label{sec.3}
\begin{figure}[htb] 
	\centering
	\includegraphics[width=3.0 in]{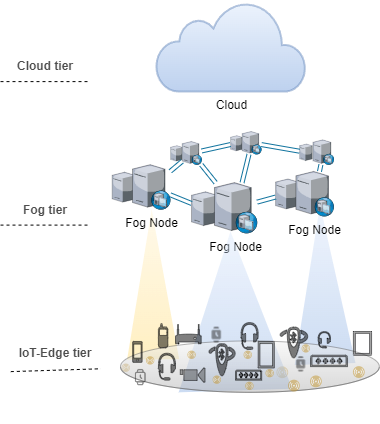}
	\caption{IoT-Edge based Fog infrastructure }
	\label{fig:Architecture}
\end{figure}

In this section, we introduce a fog deployment model within the IoT-fog-cloud infrastructure. The studied system includes IoT sensors, devices, and fog nodes as shown in figure~\ref{fig:Architecture}.  In order to fulfill terminal IoT edge nodes requirements for high throughput and high quality of service while keeping costs low IoT-edge-based fog infrastructure has been presented in Figure \ref{fig:Architecture}. Which presents the core concepts of the model involving system architecture, coverage, and network connectivity, as well as multi-objective fog node deployment challenges. 

\begin{table}
	\centering
	\caption{mathematical notation used in model development}
    \label{tabs:notation}
	\resizebox{\linewidth}{!}{%
		\begin{tabular}{|>{\hspace{0pt}}p{0.159\linewidth}|>{\hspace{0pt}}p{0.752\linewidth}|} 
			\hline
			Symbol & Meaning \\ 
			\hline
			$n$ & Number of fog nodes \\ 
			\hline
			$m$ & Number of IoT-edge clients \\ 
			\hline
			$l$ & Location vector of fog nodes \\ 
			\hline
			$l(x_i,y_i)$ & Node location \\ 
			\hline
			$W$ & Region of interest width \\ 
			\hline
			$H$ & Region of interest height \\ 
			\hline
			$FG$ & Fog node set\\ 
			\hline
			$EN$ & IoT-edge nodes set \\ 
			\hline
			$EN_i$ & The $i$-th edge node client \\ 
			\hline
			$E$ & Set of edge links between fog nodes \\ 
			\hline
			$G_i$ & $i$-th subgraph component \\ 
			\hline
			$\vert G_i \vert$  & Size of the $i$-th subgraph component \\ 
			\hline
			$h$ & Number of subgraph components \\ 
			\hline
			$\zeta(G)$ & Network connectivity \\ 
			\hline
			$\Phi(G)$ & Terminal IoT-edge node coverage \\ 
			\hline
            $\gamma_i$ & Boolean coverage function \\ 
			\hline
			$X_i$ & The i-th particle \\ 
			\hline
			$gBest$ & The Global best solution\\ 
			\hline
			$Max_{Itr}$ & Maximum step movement\\ 
			\hline
		\end{tabular}
	}
\end{table}

\subsection{Network Model Description}

Figure~\ref{fig:Architecture} shows the IoT-Edge-based Fog infrastructure, which has three tiers. The first tiers is the IoT-Edge tier, which includes many IoT-edge terminal nodes. These nodes are linked to the second layer, which is referred to as the fog tiers and contains many fog server nodes. The third layer is the cloud layer. Computing nodes could conduct simple analytics on the data they acquire. Let the set of computing nodes within infrastructure be represented by $U=FG \cup EN$, where the fog layer consists of $n$ fog nodes denoted by $FG=\{FG_{1},FG_{2},FG_{3}...,FG_{n}\}$ with the corresponding set of transmission ranges denoted by $R(FG_{i})=\{R_{1},R_{2},R_{3}...,R_{n}\}$, and the IoT-Edge layer has $m$ edge nodes denoted by $EN =\{E_{1},E_{2},E_{3}...,E_{m}\}$. To facilitate the problem definition notations and symbols used in the system model are listed in table~\ref{tabs:notation}.

\subsection{System model assumptions}


We investigate a deployment scenario where both fog nodes and IoT-edge nodes are considered static. To respond to a real network deployment scenario in practice, Each fog node has a different length of communications radius, and fog nodes can communicate with each other via their radio coverage \cite{peng2020artificial}. On the other hand, edge nodes only have the essential functions for network connectivity, but do not have the function of gateways or bridges \cite{ghosh2021sega}. Hence, edge nodes must go through fog nodes to communicate with other nodes. Moreover, since the locations of fog nodes are determined on the basis of edge nodes' locations, it is necessary to know the locations of edge nodes in advance \cite{gilbert2021evolutionary}. Briefly, to ensure network connectivity and coverage,  the following assumptions have been established:

 \begin{itemize}
	\item The fog nodes and IoT-edge nodes are deployed uniformly and randomly over the entire network and are considered static inside the region of interest.
	\item Each fog node in the region of interest has access to the cloud center via cellular networks; the terminal nodes' placements are considered fixed.
	\item To adapt to the heterogeneity of terminal nodes in practice, each fog node $fg_{i}$ is assumed to have a different transmission range $R_{fg_i}$.
    \item A fog node can only be attached to a certain number of edge nodes.
\end{itemize}

Note that Euclidean distance is an important metric for determining network accessibility and connectivity. Therefore, two conditions must be satisfied:

\begin{enumerate}
  \item An Iot-edge node $EN_{i}$ is considered connected, if and only if it is covered by at least a fog node $fg_{i}$, as shown in eq(\ref{eq_range}: 
  
\begin{equation}
\sqrt{\left(x_i-x_j\right)^2+\left(y_i-y_j\right)^2} \leq R_i
\label{eq_range}
\end{equation}.

  \item Two fog nodes, $fg_{i}$, and $fg_{j}$ are regarded as connected. if and only if they are in the same transmission range, as shown in eq(\ref{eq_minrange}: 

\begin{equation}
\sqrt{\left(x_i-x_j\right)^2+\left(y_i-y_j\right)^2} \leq \min \left(R_i, R_j\right)
\label{eq_minrange}
\end{equation}

\end{enumerate}

\begin{figure*}[!htbp]
	\centering
	\begin{subfigure}[b]{0.49\textwidth}
		\centering
		\includegraphics[width=\textwidth]{ 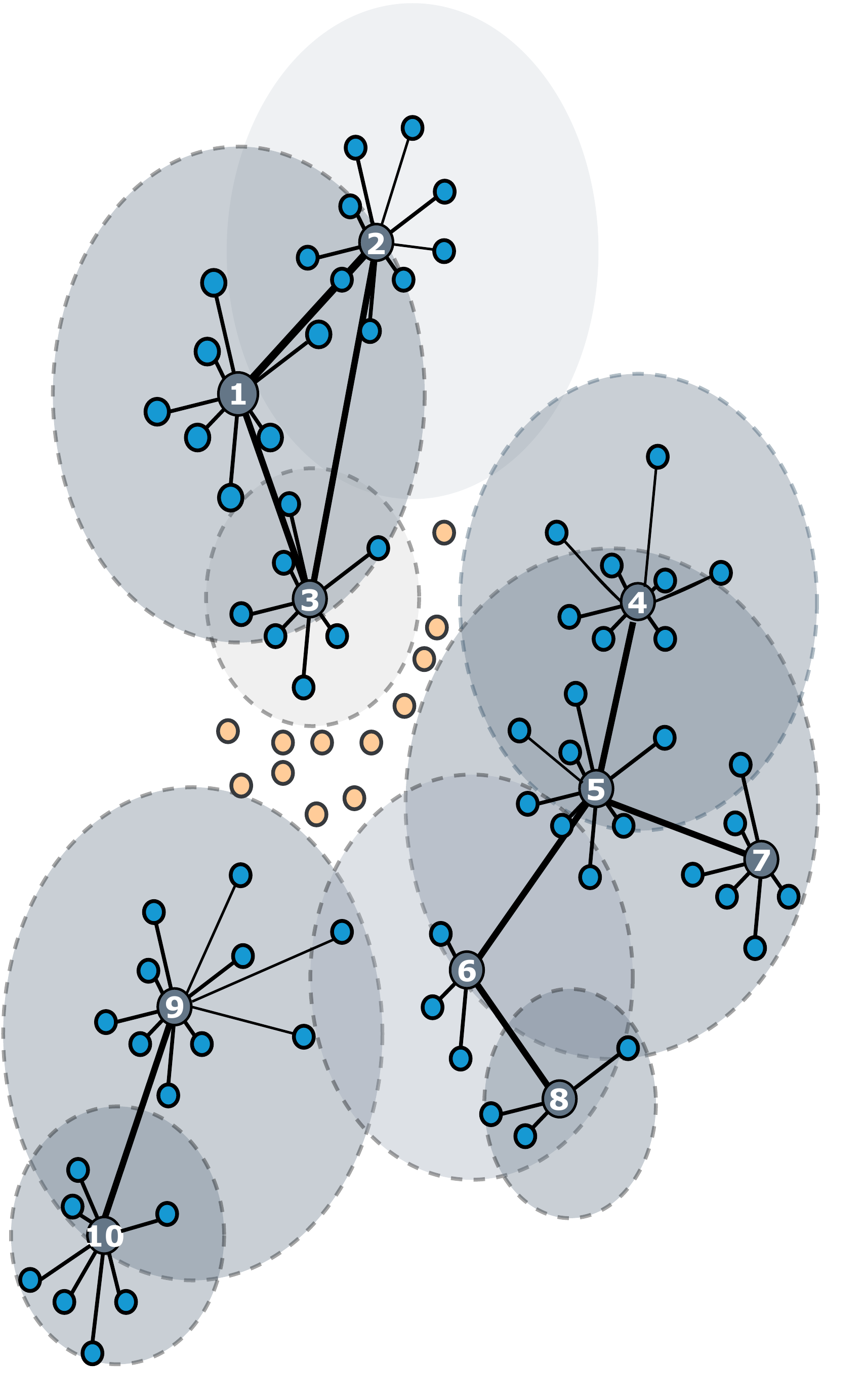}
		\caption{Fragmented network topology}
		\label{fig:test1}
	\end{subfigure}
	\hfill
	\begin{subfigure}[b]{0.49\textwidth}
		\centering
		\includegraphics[width=\textwidth]{ 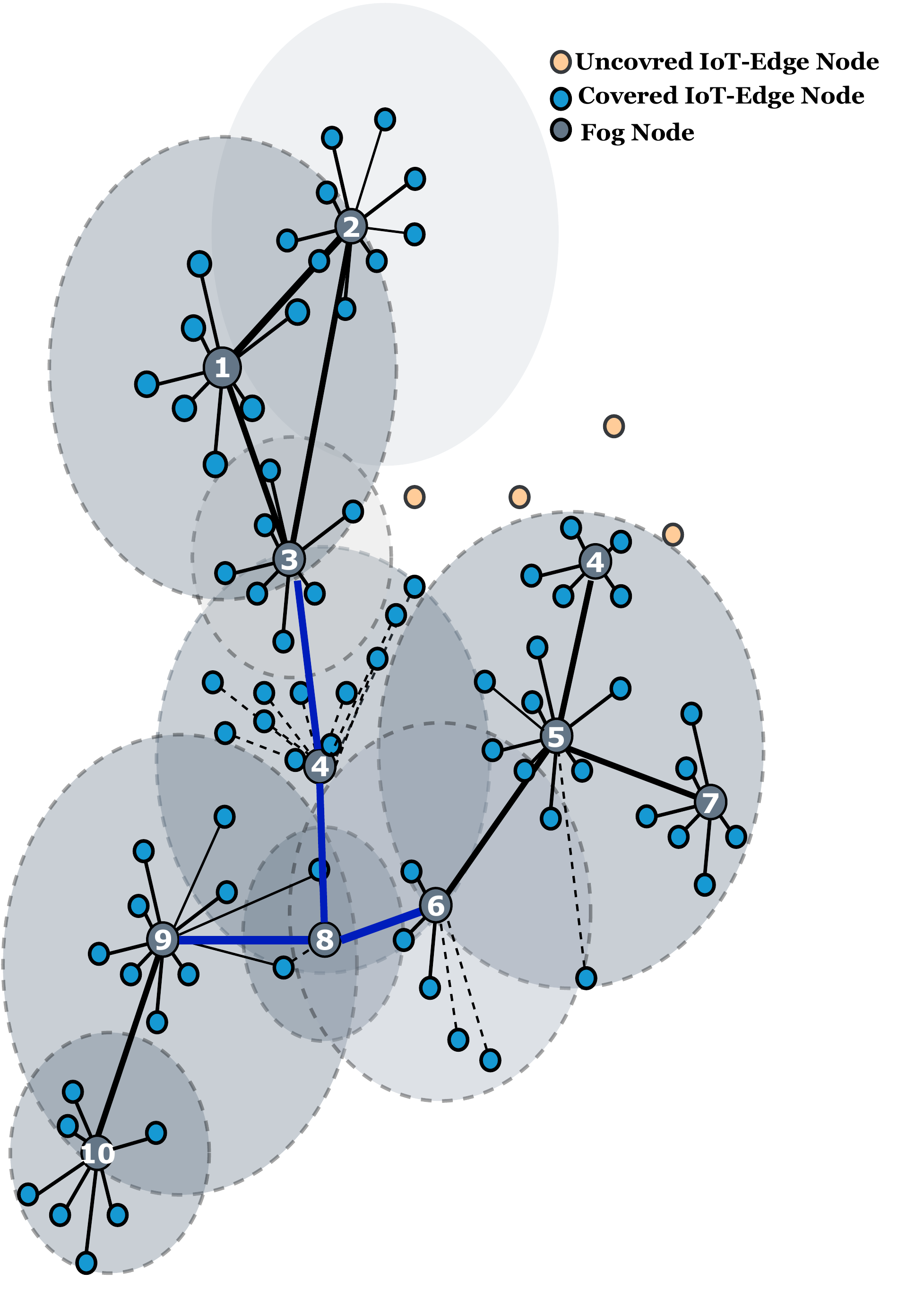}
		\caption{Near-optimal network topology}
		\label{fig:test2}
	\end{subfigure}
	\caption{Fog connectivity and  Iot-Edge nodes coverage}
	\label{fig:fig01}
\end{figure*}

Figure~\ref{fig:fig01} demonstrates an instance of the introduced problem, in this example nodes consist of two types: fog and IoT-edge nodes, which consists of 10 fog nodes n = 10 and 80 IoT-edge devices m = 80, located within the deployment area, each fog node has a varied communication range. If two fog nodes are in the communication range of each other, they will be linked via a dark link (e.g., see the dark link between fog nodes 1 and 2). If an IoT-edge node is located within the communication range of a fog node, it will be connected by a thin black link to the nearest fog. Furthermore, the topology graph includes three sub-graph components, the largest of which is 5 (i.e., $\zeta(G)=5$), and 68 IoT-edge devices are covered (i.e., $\Phi(G)==68$). If we shift certain fog nodes toward the most congested part of the IoT-edge device, as illustrated in Figure ~\ref{fig:test1}, practically all of the sub-graphs will be combined into a single giant graph with size 10 (i.e., $\Phi(G)$ = 10), and 76 IoT-edge devices will be covered (i.e., $\zeta(G)$= 76). Therefore, both metrics: edge coverage and network connectivity can be improved by changing the locations of some fog nodes.
 
\subsection{Problem Formulation}

We focus on determining near-optimal fog node placement by placing fog nodes in appropriate positions within the region of interest while ensuring connectivity and maximizing coverage. The network coverage refers to the number of covered edge nodes while the connectivity refers to connected fog nodes within the network. To address this issue, it is not practical to analyze the whole network, which is hard or even impossible to provide a connected network that covers all edge nodes due to the network dis-connectivity which can be treated separately instead. Therefore, in our study, we target the greatest sub-network i.e. largely connected sub-network.  By assuming L is the initially given location set of fog nodes $L=\left\{L\left(x_1,y_1\right), L\left(x_2, y_2\right),\ldots, L\left(x_n, y_n\right)\right\}$. 

Our objective is to update the fog nodes' locations in such a way that the covered edge nodes and the size of the greatest sub-network of connectivity are maximized. We observe that these two targets are in dispute, which means that a large sub-component network does not necessarily indicate a wide coverage of edge nodes. We analyze our scenario problem with a network of n fog nodes and m edge devices deployed in a 2D area. We represent an undirected topology graph $G=(V,E)$, where $V= FG\cup EN$. $V$ refers to network nodes which include a set of fog nodes denoted by $FG$ and a set of edge nodes denoted by $EN$, while $E$ defines the edge node's connectivities. Two fog nodes are considered connected if an edge $(fg_{i}, fg_{j}) \in E$ exists and $R_fg_i\cap R_fg_j \neq \emptyset;$ an edge node $EN_{j} \in E$ are considered covered if and only if fog node $fg_{i} \in F$, $L_{EN_j} \in R_{j}$, and edge-fog $edge(en_{j}, fg_{i}) \in E$ exists.


As mentioned earlier, It is challenging to find a fully connected and covered graph based on analyzing the entire network due to network disconnectivity, Therefore, targeting a large sub-network could be a solution to improve network connectivity. We pointed out that the corresponding graph G of the target network may not be connected, i.e., G may consist of several sub-graph components. However, note that maximizing network connectivity of fog nodes may not be able to cover all edge nodes. In this situation, we aim to render the size of the largest sub-graph component as large as possible to maximize the connectivity of the network. 

Consider that graph G contains  $h$ subgraphs $G_1, \dots, G_h$  in $G$, i.e., $G= G_1{\bigcup G}_2\bigcup\ G_h$ , and $G_i\cap G_j=\emptyset$ ; for $i,j \in {1, \dots, h}$.  In order to analyze and evaluate the performance of the topology graph G, the following metrics network coverage and connectivity are considered to be optimized. 

Network connectivity is defined as the size of the largest connected fog nodes subgraph. where G is a graph and $G_h$ is a subset of G. The network connectivity is calculated as follows:

\begin{equation}\label{eqn:connectivity}
\zeta(G)=\max\limits_{i\in\left\{1,\dots,h\right\}}|G_i|
\end{equation}

The network coverage function $\gamma_i$  of an edge node i, relative to a fog node is defined  by a binary value as follows:

$$
\gamma_i= \begin{cases}1 & \text { if the client } i \text { is covered by at least one fog node } \\ 0 & \text { Otherwise }\end{cases}
$$

Where 

\begin{equation}\label{eqn:coverage}
\Phi(G)=\sum_{i=0}^m \gamma_i
\end{equation}

\subsection{Objective function}\label{section:Objectivefunction}

To assess the performance of the introduced topology graph, we consider two objectives to optimize: network connectivity by maximizing the size of the greatest subgraph component $\zeta(G)$ and the node edge coverage $\Phi(G)$, which are defined by equation. \ref{eqn:connectivity} and \ref{eqn:coverage}, respectively. Therefore, we use the weighted sum method that transforms the multi-objective problem into a scalar problem by summing each objective pre-multiplied by a user-provided weight.
Our aggregated fitness function $f(X)$ is defined as follows:
\begin{equation}\label{eqn:objective_function}
f(X)=\omega.\frac{\zeta(G)}{n}\ +\ (1\ -\ \omega)\ .\frac{\ \Phi(G)\ }{m}
\end{equation}

where $\omega$ is a weighting coefficient between zero and one that describes the ratio in which the objectives are prioritized.


\section{Marine Predators Approach}
\label{sec.4}
The MPA algorithm \cite{FARAMARZI2020113377} was recently proposed to mimic the behavior of marine predators in pursuit of their food, where the predator trades off the Brown and Levy strategy by using the speed ratio between predator and prey, The major motivation for the MPA algorithm is the extensive foraging strategy of ocean predators, namely Lévy and Brownian motions ~\cite{Humphries2010}, and encounter optimal rate policy in biological interactions between predator and prey. MPA adheres to the criteria that naturally regulate optimal foraging strategy and encounter rate policy in marine habitats~\cite{Bartumeus2002}. According to the MPA definition, the MPA-based optimization algorithm is presented as follows:

At the initiation phase, a number of the prey initially dispersed uniformly around the search area of the optimization process. Equation \ref{eqn:initiation} illustrates its mathematical model.
\begin{equation}\label{eqn:initiation}
X_0=X_{\min }+\text { rand } \times\left(X_{\max }-X_{\min }\right)
\end{equation}

$X_{max}$ and $X_{min}$, respectively, define the search space's boundaries, where $X_{max}$ is the upper bound and $X_{min}$ is the lower bound, rand is a variable generated randomly between 0 and 1. 

During the optimization phase, each distributed solution is evaluated using the objective function, and the solution with the best objective value is chosen as the top predator. This later is used to create a matrix called Elite, also referred to as E, which is based on the notion of the survival of the fittest. The elite matrix is demonstrated as follows :

\begin{equation}
\text { Elite }=\left[\begin{array}{cccc}
X_{1.1}^1 & X_{1.2}^1 & \ldots & X_{1 . d}^1 \\
X_{2.1}^1 & X_{2.2}^1 & \ldots & X_{2 . d}^1 \\
\vdots & \vdots & \vdots & \vdots \\
X_{n .1}^1 & X_{n .2}^1 & \ldots & X_{n . d}^1
\end{array}\right]_{\mathrm{nxd}}
\end{equation}

The variables X1, n, and d, respectively,  represent the top predator vector which is repeated N times to construct Elite matrix, the number of search agents within the population, and the number of dimensions. The search agents predator and prey both their tasks are based on space exploration to locate food. The Elite matrix updates each iteration according to the better predator, in comparison to the top predator in the previous iteration.  Typically, the construction of the prey occurs during the initiation phase, when the predators change their positions. Because of this, a matrix called prey has the same dimensions as the Elite matrix is defined. The fittest of the original prey is utilized to build the Elite matrix.  This matrix is formulated as follows: 

\begin{equation}
\text { Prey }=\left[\begin{array}{cccc}
X_{1.1} & X_{1.2} & \ldots & X_{1 . d} \\
X_{2.1} & X_{2.2} & \ldots & X_{2 . d} \\
\vdots & \vdots & \vdots & \vdots \\
X_{n .1} & X_{n .2} & \ldots & X_{n . d}
\end{array}\right]_{n \times x d}
\end{equation}

$X_{i.j}$ denotes the $j_{th}$ dimension of the $i_{th}$ prey. The two matrices described previously perform a significant role in the overall optimization problem. Based on the predator and prey movement patterns. The MPA optimization process is separated into the following stages:

\subsubsection{HIGH-VELOCITY RATIO (Stage1)}

At this stage, the prey is traveling faster than the predator in relation to its speed when exploration is important. This stage usually occurs early in the initial stage of the iteration process, Therefore this stage is defined as an exploration level, in which the prey starts to explore the search space using the Brownian strategy to identify the prospective areas that could contain the ideal solution. Consequently, mathematical formalism can be defined as:

\begin{equation}
\text { While } t<\frac{1}{3} T_{\max } i=1,2, \ldots, n
\end{equation}

\begin{equation}
\left\{\begin{array}{l}
\text { stepsize }_i=\left(R_B \otimes\left(\text { Elite }_i-R_B \otimes \text { pre }_i\right)\right) \\
\text { prey }_i=\text { prey }_i+P \times R \otimes \text { stepsize }_i
\end{array}\right.
\end{equation}

where $t$ represents the current iteration count, $T_{max}$ represents the algorithm's maximum cycle count, stepsize represents the motion scale factor, $R_B$ represents a Brownian walk random vector that follows the normal distribution, $Elite_i$ represents the elite matrix formed by the top predator, $prey_i$ represents the prey matrix, $\otimes$ stands for elementwise multiplication, $R$ is a random variable with a range of [0, 1], whereas $P$ is a fixed value set to 0.5.

\subsubsection{RELATIVE UNIT VELOCITY(Stage2)}
This stage is based on  exploration and exploitation \cite{hussain2019exploration}. After completing the first third of the exploration process, both predator and prey explore the search space at the same velocity in order to search for their own food. This is due to the proximity of the potential locations that may contain the ideal solution, therefore this stage is considered as an intermediate stage of optimization. In summary, in this stage, the  exploration process will be progressively converted into the exploitation process. Specifically, the predators will be responsible for the exploration, while the prey for the exploitation. To model this stage mathematically, the population is divided into two parts: the first  one is exploited by prey using a Lévy walk, while the second is explored by the predator using Brownian motion.

\begin{equation}
\text { While } \frac{1}{3} T_{\max }<t<\frac{2}{3} T_{\max } i=1,2, \ldots, n / 2
\end{equation}

\begin{equation}
\left\{\begin{array}{l}
\text { stepsize }_i=\left(R_L \otimes\left(\text { Elite }_i-R_L \otimes \text { prey }_i\right)\right) \\
\text { prey }_i=\text { prey }_i+P \times R \otimes \text { stepsize }_i
\end{array}\right.
\end{equation}

\begin{equation}
\text { While } \frac{1}{3} T_{\max }<t<\frac{2}{3} T_{\max } i=n / 2, \ldots, n
\end{equation}

\begin{equation}
\left\{\begin{array}{l}
\text { step size }_i=\left(R_B \otimes\left(R_B \otimes \text { Elite }_i-\text { prey }_i\right)\right) \\
\text { prey }_i=\text { Elite }_i+P \times \mathrm{CF} \otimes \text { step size }_i
\end{array}\right.
\end{equation}

Where $R_L$ is the random vector facilitating Lévy's mobility and $CF$ is an adaptive variable  determined by equation \ref{eqn:random}; works on  controlling the predator's path.
\begin{equation} \label{eqn:random}
C F=\left(1-\frac{t}{T_{\max }}\right)^{\frac{2 t}{T_{\max }}}
\end{equation}

\subsubsection{SLOW VELOCITY RATIO (Stage3)}
In the last third of the maximum iteration  of the optimization process, the predator starts to approach the prey following Lévy’s motion. This stage can be modeled as:
\begin{equation}
\text { While } t>\frac{2}{3} T_{\max }, i=1,2, \ldots, n
\end{equation}

\begin{equation}
\left\{\begin{array}{l}
\text { stepsize }_i=\left(R_L \otimes\left(R_L \otimes \text { Elite }_i-\text { prey }_i\right)\right) \\
\text { prey }_i=\text { Elite }_i+P \times C F \otimes \text { stepsize }_i
\end{array}\right.
\end{equation}

The flexibility of MPA to mimic the habits of predators to maximize the chance of escape from local optima is another benefit \cite{zhong2021mompa}. This effect is that predator movement would be impacted by environmental factors including eddy formation and fish aggregation devices (FADs). Its mathematical model is shown in equation \ref{eqn:FADs}:

\begin{equation}\label{eqn:FADs}
\text { prey }_i= 
\begin{cases}
\text { if } r \leq P_f  \\ 
\text { prey }_i+C F\left[X_{\min } +R \otimes\left(X_{\max }+X_{\min }\right)\right] \otimes U 
\\ \text { if } r>P_f  \\
\text { prey }_i+\left[P_f(1-r)+r\right] \times\left( 
\text { prey }_{r 1}-\text { prey }_{r 2}\right) 

\end{cases}
\end{equation}

    


   
          

         
          

$P_f$ indicates the likelihood that FADs would affect the optimization process; $U$ denotes a binary vector containing 0 and 1 value; $r$ a distinct value falling between [0, 1]; and the values $r1$ and $r2$, produced at arbitrary from the overall population range which stand for the random indices of the prey matrix.

MPA may conserve data by preserving the prey's prior location. The fitness values of each current solution and each obsolete solution are compared once the existing solutions have been updated, and if the fitness of the previous solution is higher than that of the latter, the two are exchanged. The major steps of the applied marine predators optimization method (MFO) are outlined in algorithm \ref{algo:algo1}.

\section{EXPERIMENTAL AND RESULTS}
\label{sec.5}
In this section, we discuss the achieved results using the proposed architecture. A three-tier IoT-edge fog-based architecture with IoT-edge devices located at the first tier that is focused on producing data from various IoT-edge devices. The second comprises fog servers, and the last tier includes a resource-rich cloud. This architecture facilitates the deployment and the usage of resources while specifying the responsibilities of each tier. In order to assess the performance of the presented MPA algorithm. First, we describe the experience data and environment. Then, we conduct various scenarios under different settings and compare them to different algorithms.

\subsection{Experimental setup}


In the experimental setting, we implement a practical application using  Matlab 2019Ra. This permits the calculations to be conducted as fast as possible. Based on the pseudo-code presented in algorithm \ref{algo:algo1}, we conducted our experiment in a rectangular area of 1000m × 1000m with normal/uniform distribution to fog nodes using an AMD Ryzen 7 5700U with 8 cores with a 4.3 GHz clock speed, 8GB of memory, and a Windows 10 environment. The experimental setup enables us to repeat the experiment under various conditions and restrictions allowing us to conduct the experiment in the control environment with a desired set of settings. Table ~\ref{tabs:NetworkParameters} provides a list of parameters and their corresponding values, which were derived and established by various initial experiments.

\begin{table}[!htbp]
	\centering
	\caption{Network Parameters}
	\label{tabs:NetworkParameters}

		\begin{tabular}{@{}lll@{}} 
			\toprule
			Parameter & Value & Initial value \\ 
			\midrule
			Fog nodes density & {[}10, 120] & 45 \\ 
			
			Edge nodes density & {[}30, 200] & 120 \\ 
			
			Communication range & {[}90, 200] & 100m \\ 
			
			Area width & 1000m & 1000m \\ 
			
			Area height & 1000m & 1000m \\ 
			\bottomrule
			
		\end{tabular}
	
\end{table}

\subsection{Experimental Results}
We have conducted various to evaluate the connectivity and coverage of the proposed algorithm compared to the following baselines: harris hawks optimization (HHO) \cite{HHO2019}, particle swarm optimization (PSO) \cite{PSO2017} and sine cosine algorithm (SCA) \cite{sca2016} algorithms. Due to the fact that our target function involves two competing priorities and that the network may experience inadequate connectivity and coverage as illustrated in Figures~\ref{fig:test1},~\ref{fig:test2}. Therefore, the conducted tests have been carried out under different network parameters to establish the best weight $\omega$ of the defined objective function for comparing the performance of the proposed algorithm to other methods.


\begin{algorithm}[!htbp]
	\DontPrintSemicolon
	\SetAlgoLined
    
	\KwInput{Search agents (Prey) population $X_i(i=1,2,3...,n)$, }
	\KwOutput{Optimal fitness value of fog node deployement $fit_{best}$}

	\While {$iter < Max_{Itr}$ } {
        Calculate the fitness value of each $p y_i \mid i=0,1,2,3,.....N$\;
        Build the elite Matrix E and accomplish memory saving\;

        \If{$ iter< Max_{Itr}/3$} {
          Update prey based on Eq. (2)\;
		 } 
   
        \If{$Max_{Itr}/3 < iter <2.Max_{Itr}/3$} {
          For the first half of the populations $(i=1, \ldots, n / 2)$ \;
          Update prey based on Eq. (3) \;
          
          For the other half of the populations $(i=n / 2, \ldots, n)$ \;
          Update prey based on Eq. (4)\;
		 } 

        \If{$iter >2.Max_{Itr}/3$} {
         
          Update prey based on Eq. (6)\;
		 } 
          
        Accomplish memory saving and Elite update\;
        Applying FADs effect and update based on Eq. (7)\;
        \text { Iter=Iter }+1 \\
	}    
	\caption{Pseudo code  of Marine Predator Optimization algorithm}
	\label{algo:algo1}
\end{algorithm}
\subsection{Fitness Function Study Under Different $\omega$ Values }

\begin{figure*}
	\centering
	\includegraphics[width=\textwidth]{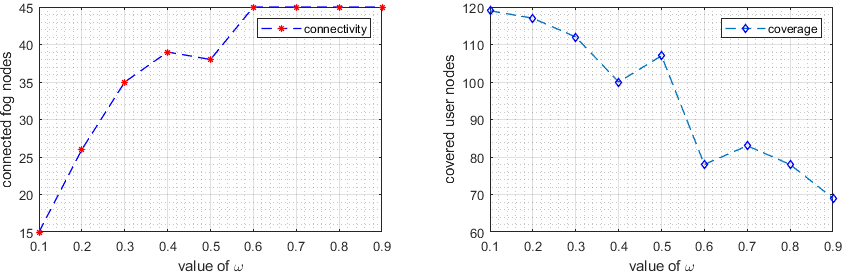}
	\caption{Fitness Function Study Under Different $\omega$ Values}
	\label{fig:omega}
\end{figure*}

To determine an appropriate weight-coefficient $\omega$, we analyze the user coverage and fog nodes connectivity simultaneously with different values of $\omega$, and we employ the convergence curves which are considered the most critical analysis tool to understand the algorithm's behaviors while developing an optimal solution using metaheuristics algorithms. Therefore, to identify the optimum weight-coefficient $\omega$, several iterations were conducted until the termination criteria in equation~\ref{eqn:objective_function} are satisfied. We execute the experiments 10 times, each of which has 1000 iterations. And, each experiment was associated with different values of $\omega$. In each experiment, 1000 instances are employed to determine the effect of changing the weight-coefficient $\omega$ value. We observed in that when using a larger value of $\omega$, the user coverage is low which means fog nodes have a tendency to crowd in a small area, leaving many IoT-edge nodes uncovered, whereas when using a smaller value, the user coverage is high which mean the fog nodes are dispersed throughout the area and can cover the majority of IoT-edge nodes independently but cannot make up a larger component(i,e, High connectivity). Hence we conclude, to some measure, that the weight-coefficient $\omega$ defines the tradeoff between our two concerned objectives ($\zeta$ and $\Phi$). Therefore, we determine the weight coefficient $\omega$, which is most likely to meet the two relevant criteria instantaneously. After numerous tests, figure ~\ref{fig:omega} shows that the obtained results with the value of $\omega = 0.5$ give an excellent tradeoff.

Figures \ref{fig:convergance} show obtained results of studied systems with respect to different network configuration values. As noticed at the beginning of an experiment, all algorithms undergo a significant change; nevertheless, this change gradually decreases as the iteration progresses. On the other hand, the  HHO, PSO and SCA algorithm has trouble with premature convergence, which makes them more likely to get stuck in a local optimum than a global one. A crucial advantage of an algorithm's overall performance is its velocity of progression. As demonstrated in Figures~\ref{fig:convergance}, after a certain number of iterations, the proposed algorithm showed excellent potential and faster convergence than other algorithms while also achieving an optimal fitness function value. Hence, the results show that the proposed algorithm outperforms other algorithms in solving fog deployment problems for different aggregation values of $\omega$.


\begin{center}
	\includegraphics[width=3.5in]{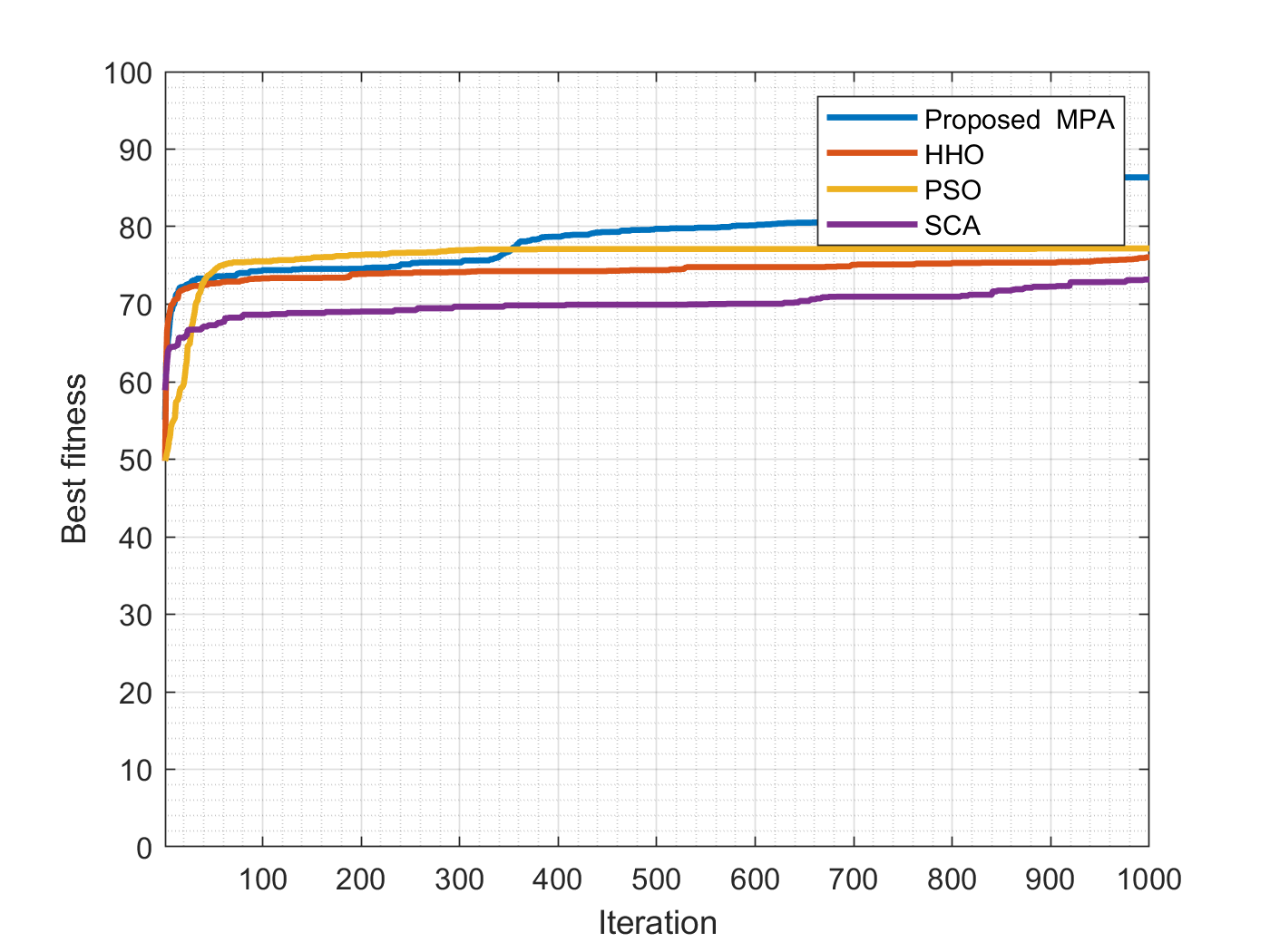}
	\captionof{figure}{Convergence curve of studied algorithms}
	\label{fig:convergance}
\end{center}

\subsubsection{Comparison of algorithms convergence}

\begin{figure*}[!htbp]
	\centering
	\begin{subfigure}[b]{0.45\textwidth}
		\centering
		\includegraphics[width=\textwidth, height=5cm]{ 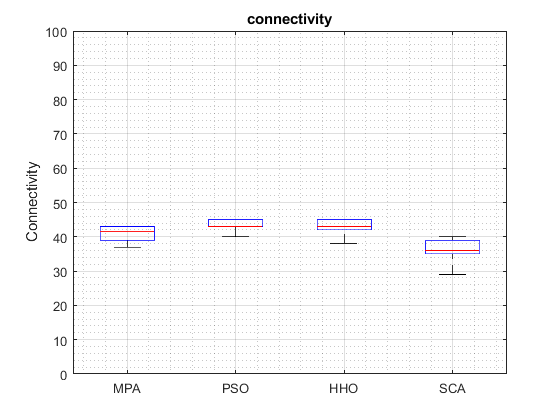}
		\caption{Network connectivity}
		\label{fig:connectivity}
	\end{subfigure}
	\hfill
	\begin{subfigure}[b]{0.45\textwidth}
		\centering
		\includegraphics[width=\textwidth, height=5cm]{ 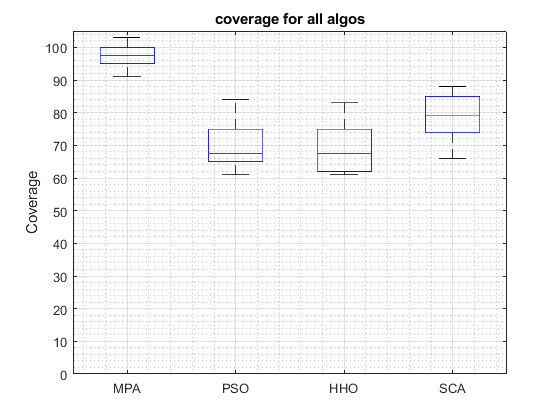}
		\caption{Iot-edge nodes coverage}
		\label{fig:coverage}
	\end{subfigure}
	\caption{Comparison of MPA,PSO, HHO, and SCA}
	\label{fig:connectivityAndcoverage}
\end{figure*}

To assess how well the proposed algorithm performs in comparison to PSO, HHO, and SCA algorithms for the proposed objective function $f$ as given in equation \ref{eqn:objective_function}. Both the network and user coverage are examined. Figure \ref{fig:connectivityAndcoverage} shows how the proposed algorithm can cover more IoT-edge nodes with fewer fog nodes, whereas in Figure \ref{fig:connectivity} 40\% percent of connected fog nodes can cover more than 90\% of the target population as shown in Figure~\ref{fig:coverage}, while with PSO, HHO, and SCA only connect (45,45,41)\% of fog servers with covering (67,67,79)\% of terminal edge devices, respectively. Hence, the results showed that our presented MPA algorithm outperforms better than other methods.

\subsection{EVALUATION CRITERION}

In order to have a better understanding of the adjustments that have been made to the optimization objectives, we have evaluated the performance of the proposed marine predator's optimization algorithm against the previous methods, in terms of the following evaluation metrics, namely; Network connectivity ($\Phi$), user coverage ($\zeta$), and objective function value $f$, we assess our experiment by varying the number of edge nodes, the number of fog nodes, and the fog nodes' transmission range. 


\begin{figure*}[!htbp]
	\centering
	\begin{subfigure}[b]{0.45\textwidth}
		\centering
		\includegraphics[width=\textwidth, height=5cm]{ 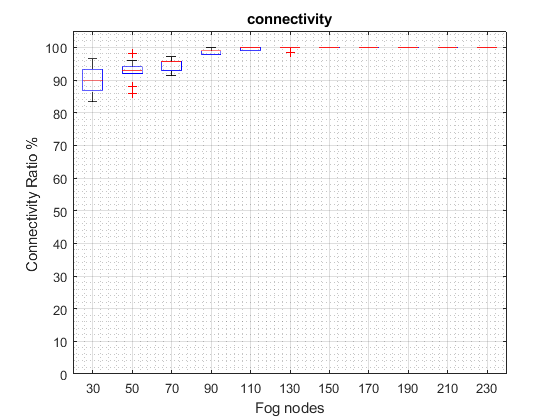}
		\caption{Network connectivity}
		\label{fig:fig21a}
	\end{subfigure}
	\hfill
	\begin{subfigure}[b]{0.45\textwidth}
		\centering
		\includegraphics[width=\textwidth, height=5cm]{ 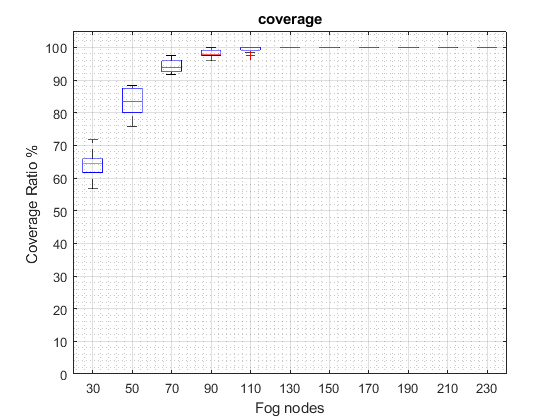}
		\caption{Iot-edge nodes coverage}
		\label{fig:fig21b}
	\end{subfigure}
	\caption{Effect of fog nodes density on coverage and connectivity}
	\label{fig:fig21}
\end{figure*}

\begin{figure*}[!htbp]
	\centering
	\begin{subfigure}[b]{0.45\textwidth}
		\centering
		\includegraphics[width=\textwidth, height=5cm]{ 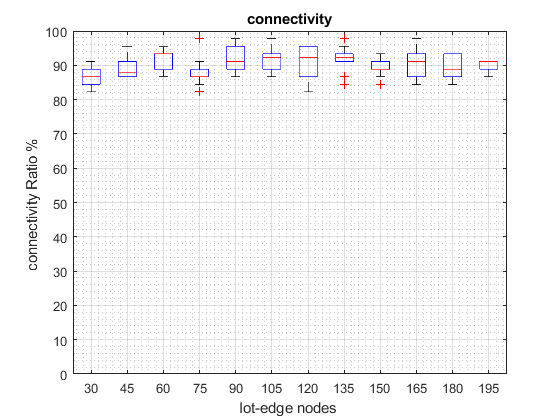}
		\caption{Network connectivity}
		\label{fig:fig22a}
	\end{subfigure}
	\hfill
	\begin{subfigure}[b]{0.45\textwidth}
		\centering
		\includegraphics[width=\textwidth, height=5cm]{ 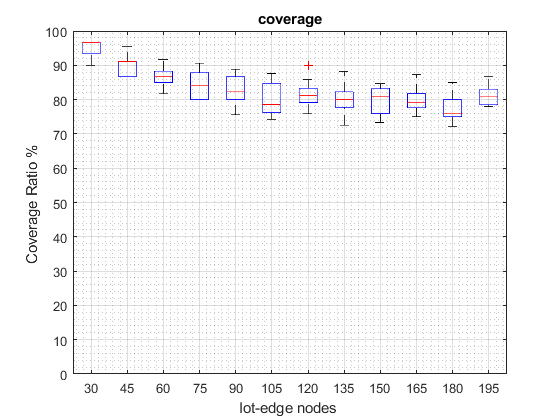}
		\caption{Iot-edge nodes coverage}
		\label{fig:fig22b}
	\end{subfigure}
	\caption{Effect of IoT-edge nodes density on coverage and connectivity}
	\label{fig:fig22}
\end{figure*}

\begin{figure*}[!htbp]
	\centering
	\begin{subfigure}[b]{0.45\textwidth}
		\centering
		\includegraphics[width=\textwidth, height=5cm]{ 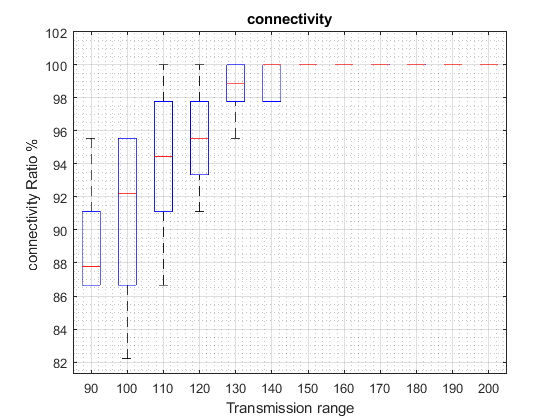}
		\caption{Network connectivity}
		\label{fig:fig23a}
	\end{subfigure}
	\hfill
	\begin{subfigure}[b]{0.45\textwidth}
		\centering
		\includegraphics[width=\textwidth, height=5cm]{ 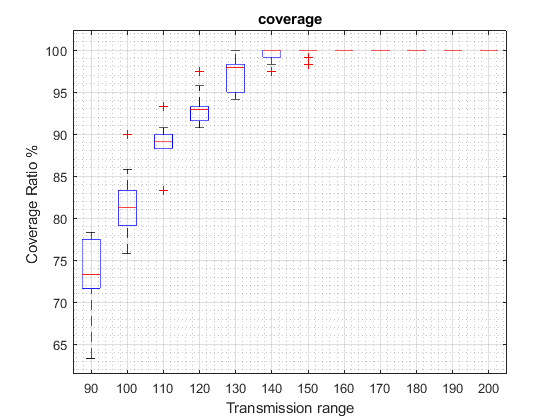}
		\caption{Iot-edge nodes coverage}
		\label{fig:fig23b}
	\end{subfigure}
	\caption{Effect of the transmission range on coverage and connectivity}
	\label{fig:fig23}
\end{figure*}

\subsubsection{The impact of network  density}

To analyze the network coverage and connectivity, we establish a fog network by deploying a variety of network sizes (N), varying from 10 to 120 fog nodes as shown in Figures~\ref{fig:fig21},~\ref{fig:fig22} and~\ref{fig:fig23}. As can be seen in Figure~\ref{fig:fig21a}, the number of connected fog nodes increases as additional fog nodes are added to the network which reflects the network connectivity. As predicted, increasing the number of fog nodes increases network connectivity. This is due to the fact that by adding additional fog nodes to the network the isolated network connects to the rest of the network. As a direct consequence of this, a wider network will be formed. While, in terms of coverage, Figure~\ref{fig:fig21b} displays the relationship between the number of fog nodes as well as the total number of covered IoT-edge nodes. The results show that our proposed algorithm finds optimum coverage, with more IoT-edge nodes covered as predicted. 

\subsubsection{The impact of edge node density}

Figure~\ref{fig:fig22} shows the network coverage and connectivity results by varying the total number of IoT-edge nodes from 30 to 200. From figure~\ref{fig:fig22a}, it is clear that the number of connected fog nodes is always somewhere between 70\% and 90\%. This is due to the fact that our  algorithm makes an effort to connect all of the accessible fog nodes in an attempt to encompass additional edge nodes. Furthermore, figure~\ref{fig:fig22b} shows the network's overall coverage which remains stable between 70\% and 93\%. We observed that when the number of IoT-edge nodes increased, the percentage of coverage and connectivity remained consistent. This is because the IoT-edge node distribution is uniform, which implies that the likelihood of an IoT node falling into a covered or uncovered region is the same.

\subsubsection{The impact of communication range}
Figure~\ref{fig:fig23}, shows the communication range adjustments' impact on the network coverage and connectivity performance across multiple experimental runs. By adjusting the range of the communication from 90 meters all the way up to 200 meters respectively. The result in Figure~\ref{fig:fig23a} shows that once the communication range exceeds 140 meters, all fog nodes start attempting to communicate with each other. This happens because almost isolated sub-networks connect to the rest of the network forming a single large giant network. Figure~\ref{fig:fig23b} shows the impact of communication range changes on IoT-edge nodes coverage. Results proved that the number of covered edge nodes grows proportionally with the transmission range. According to Figure~\ref{fig:fig23b}, when the communication range goes over and above 140 meters, almost terminal nodes of the network are covered. Hence, 140m is considered a crucial communication range.



\begin{table}[!htbp]
	\centering
	\caption{\label{tab:table-f-fog}Coverage, connectivity, and fitness value vs Fog node density}
	\label{tab:table-f-mrs}
	\resizebox{\linewidth}{!}{
		\begin{tabular}{c c c c} 
  \hline
			Fog & Connectivity\%) & Coverage (\%) & function \\ 
			\midrule
			30	&	90,00	&	64,25	&	77,13	 \\
			50	&	92,60	&	83,50	&	88,05	 \\
			70	&	94,86	&	94,00	&	94,43	 \\
			90	&	98,67	&	98,00	&	98,33	 \\
			110	&	99,73	&	99,50	&	99,61	 \\
			130	&	99,85	&	100,00	&	99,92	 \\
			150	&	100,00	&	100,00	&	100,00	 \\
			170	&	100,00	&	100,00	&	100,00	 \\
			190	&	100,00	&	100,00	&	100,00	 \\
			210	&	100,00	&	100,00	&	100,00	 \\
			230	&	100,00	&	100,00	&	100,00	 \\

			\bottomrule
		\end{tabular}
	}
\end{table}

\begin{table}[!htbp]
	\centering
	\caption{\label{tab:table-f-node}Coverage, connectivity, and fitness value vs IoT-edge node density}
	\label{tab:table-f-mrs}
	\resizebox{\linewidth}{!}{
		\begin{tabular}{c c c c} 
			\hline
			IoT-edge & Connectivity(\%) & Coverage (\%) & Function \\ 
			\midrule
			30	&	86,7	&	95,0	&	90,83	 \\
            45	&	89,1	&	90,4	&	89,78	 \\
            60	&	91,8	&	86,8	&	89,31	 \\
            75	&	88,0	&	84,5	&	86,27	 \\
            90	&	91,8	&	82,7	&	87,22	 \\
           105	&	92,2	&	80,3	&	86,25	 \\
           120	&	90,9	&	81,8	&	86,36	 \\
           135	&	91,8	&	80,4	&	86,07	 \\
           150	&	89,6	&	79,8	&	84,68	 \\
           165	&	90,9	&	79,7	&	85,29	 \\
           180	&	89,3	&	77,2	&	83,25	 \\
           195	&	90,0	&	81,1	&	85,54	 \\   
			\bottomrule
		\end{tabular}
	}
\end{table}

\subsubsection{Experimental objective function impact}
This part shows the impact of user coverage and network connectivity suggested in the objective function~\ref{eqn:objective_function} in the performance of the proposed algorithm. Using the same experimental parameters as before, Table ~\ref{tab:table-f-fog} and Table ~\ref{tab:table-f-node} outline accurate numerical results of the objective function obtained by the proposed algorithm. Results in Table~\ref{tab:table-f-fog}, show that when increasing the number of fog nodes from 30 to 230, the covered IoT-edge nodes percentage and value of objective function increase correspondingly. In addition, nearly all the time, we obtained a connected network topology. However, upon adding additional fog nodes to the network, our proposed algorithm gives greater IoT-edge node coverage and the best fitness value $f$. According to the statistics that are provided in Table~\ref{tab:table-f-mrs}, when the number of fog nodes is changed from 30 to 230, both the percentage of covered edge nodes and the value of the objective function increase in a proportional manner. Quite notably, our network structure is nearly always interconnected. However, the obtained results affirmed that the proposed approach provides better network coverage and better value to the fitness function whenever additional fog nodes are added to the network. 

By varying the number of terminal nodes from 30 to 195, results in Table~\ref{tab:table-f-node} show that the percentage of user coverage and network connectivity, as well as the objective function value f almost remained practically unchanged. This occurs because node placement distribution is totally randomly affected. Since the whole network is almost connected for all IoT-edge node values as shown, it is remarkable that even with the addition of terminal nodes, the coverage value will remain stable, which will result in the same proportion of all edge nodes. Similarly, statistics analysis demonstrates that the proposed algorithm performs better than other algorithms and achieves better coverage. The samples in Table~\ref{tab:table-f-range} illustrate how the communication range of fog nodes affects both coverage and network connectivity. Simply explained, an increase in communication range will lead to an increase in coverage, which in turn will result in an improvement in the fitness value. Moreover, the fitness value that can be attained using the proposed algorithm is greater than that which can be obtained using the other methods.

\begin{table}[!htbp]
	\centering
	\caption{\label{tab:table-f-mrs}Coverage, connectivity, and fitness value vs communication rang }
	\label{tab:table-f-range}
	\resizebox{\linewidth}{!}{
		\begin{tabular}{c c c c} 
			\hline
			Range & Connectivity \%) & Coverage (\%) & function \\ 
			\midrule
			90	&	88,89	&	73,42	&	79,61	 \\
			100	&	90,89	&	81,83	&	85,46	 \\
			110	&	94,00	&	88,50	&	90,70	 \\
			120	&	95,56	&	93,17	&	94,12	 \\
			130	&	98,44	&	97,00	&	97,58	 \\
			140	&	99,33	&	99,50	&	99,43	 \\
			150	&	100,00	&	99,75	&	99,85	 \\
			160	&	100,00	&	100,00	&	100,00	 \\
			170	&	100,00	&	100,00	&	100,00	 \\
			180	&	100,00	&	100,00	&	100,00	 \\
			190	&	100,00	&	100,00	&	100,00	 \\
			200	&	100,00	&	100,00	&	100,00	 \\

			\bottomrule
		\end{tabular}
	}
\end{table}

\section{Conclusion}\label{c}
\label{sec.6}
To improve network performance and provide a high user QoS at the edge of the network, efficient data and resource placement are required due to the IoT edge device resource limitation and user needs. Therefore, in this paper, we proposed a three-tier IoT edge-fog-cloud integration architecture and utilized Meta-heuristics algorithms for a seamless resource deployment process. Results show that with optimal resource deployment network performance can be enhanced. We propose an aggregated bi-objective function to maximize the network connectivity of the network measured by the number of connected fog nodes and to maximize the number of covered IoT-edge nodes simultaneously and we proposed an effective optimization algorithm. The results indicate that the proposed algorithm provided very promising results with fast convergence and low computational complexity compared to state-of-the-art baselines. Moreover, the experimental results demonstrated the efficiency of the proposed algorithm in finding network connections. A full network connection is always detected. However, network coverage is strongly influenced by the number and distribution of network IoT-edge nodes in the operational area.

\section*{Acknowledgment}
This work was supported by The National Natural Science Foundation of China under Grant 61872038.

\bibliographystyle{IEEEtran}
\bibliography{sample}

\end{document}